\documentclass[12pt,a4paper]{article}
\usepackage{latexsym}
\usepackage[latin1]{inputenc}
\usepackage[dvips]{graphics}
\usepackage{amssymb,amstext,amsfonts,amsmath}

\begin{document}




\begin{center}
{\Large 
A statistically consistent variational  approach to the renormalized mean-field theory of the  t-J model: critical hole concentrations for a paired state \\}
\vspace{0.5cm}
Jakub J\c{e}drak$^{\ast}$ and Jozef Spa\l ek$^{\dag}$  \\  
Marian Smoluchowski Institute of Physics, Jagiellonian University, \\
Reymonta 4, 30-059 Krak\'{o}w, Poland \\
\end{center}

\begin{abstract}
Recently, Fukushima [Phys. Rev. B \textbf{78} 115105 (2008)] proposed  a systematic derivation of the Gutzwiller approximation for the t-J model. In the present paper, using this approach  we construct an effective single-particle Hamiltonian, which leads to a renormalized mean-field theory (RMFT). We also use the method proposed by us recently and based on the maximum entropy principle   (MaxEnt), which in turn, yields a consistent statistical description of the problem.
On the examples  of non-magnetic superconducting d-wave resonating valence bond  (dRVB) and normal staggered-flux (SF) solutions, we compare two selections of the Gutzwiller renormalization schemes, i.e. the one proposed by Fukushima with  that  used earlier by  Sigrist et al. [Phys. Rev. B \textbf{49}, 12 058 (1994)]. 
We also  confront the results coming from our variational solutions with the self-consistency conditions build in,  with those  of the non-variational approach based on the Bogoliubov-de Gennes   self-consistent equations.  
Combination of the present variational approach   with the new renormalization scheme (taken from Fukushima's work) provides, for $t/J =3$, an upper  critical hole concentration $x_{c} \approx 0.27$ for the disappearance of the d-wave superconductivity. 
Also, the hole concentration   $x  \approx 0.125$ is obtained for the optimal doping. These  results are in rough accordance with experimental results for high-$T_c$ superconducting cuprates. 
\end{abstract}
\textbf{PACS:}   05.30.-d, 71.10.Fd, 75.10.Jm. 



\section{Introduction}

The t-J model \cite{Spalek Oles} is   commonly regarded as a minimal model capable of describing correctly the essential physics of the cuprate high-temperature superconductors. In its simplest form, used in the present paper, it is expressed by the following  Hamiltonian
\begin{equation}
 \hat{H}_{t-J} = \hat{P}_G \big(\sum_{i,j,\sigma} t_{ij} c_{i \sigma}^{\dag} c_{j \sigma} +  \sum_{\langle i j \rangle} J_{ij}~  \mathbf{S}_{i}\cdot \mathbf{S}_{j}) \hat{P}_G.
\label{t-J exact}
\end{equation}
Here $\hat{P}_G = \prod_{i}(1-\hat{n}_{i \uparrow}\hat{n}_{i \downarrow} ) $ is a Gutzwiller projection operator, ensuring that no doubly occupied sites are present. 

Unfortunately, the  rigorous treatment of  the t-J model is limited to very special cases. This is both due to the  interaction term ($\mathbf{S}_{i}\cdot \mathbf{S}_{j}$), as well as to the presence of $\hat{P}_G$ operators. Even in the  $J=0$ limit,  Hamiltonian (\ref{t-J exact}) is   not an independent-particle one. 
To proceed further, (\ref{t-J exact})  may be treated within the  mean-field approximation. This procedure yields an  single-particle Hamiltonian $\hat{H}$, being the  mean-field analogue of $\hat{H}_{t-J} $ (\ref{t-J exact}). The resulting  effective   description is termed the \textit{renormalized mean-field theory} (RMFT) \cite{Edegger}.

The crucial point  in  the construction of RMFT is an approximate treatment of the Gutzwiller projection. In a broader perspective, this  problem is closely related to an analytic evaluation of the expectation values of   operators with respect to the following  variational state 
\begin{equation}
 | \Psi \rangle = \hat{P}_{GC} | \Psi_{0} \rangle = \prod_{i}\lambda_{i \uparrow}^{\hat{n}_{i \uparrow}} \lambda_{i \downarrow}^{\hat{n}_{i \downarrow}}(1-\hat{n}_{i \uparrow}\hat{n}_{i \downarrow}) | \Psi_{0} \rangle
\label{Correlated state}
\end{equation} 
In the above,  $\hat{P}_{G}$  was replaced by a more general Gutzwiller correlator $\hat{P}_{GC}$, differing from $\hat{P}_{G}$   by the presence of the so-called fugacity factors $\lambda_{i \sigma}$  \cite{Fukushima}. For  $\lambda_{i \sigma} = 1$  we recover $\hat{P}_{G}$.  $| \Psi_{0} \rangle$ is an uncorrelated single-particle state, which, within the framework of RMFT, is chosen as the  eigenstate of the   $\hat{H}$.  
 Explicitly, we are interested in evaluation of the expressions like 
 \begin{equation}
\langle  \hat{\mathcal{O}} \rangle_{G}  \equiv \frac{\langle \Psi | \hat{\mathcal{O}}| \Psi \rangle }{\langle \Psi | \Psi \rangle} =   \frac{\langle \Psi_0 | \hat{P}_{GC} \hat{\mathcal{O}} \hat{P}_{GC}| \Psi_0 \rangle }{\langle \Psi_0 | \hat{P}^{2}_{GC} | \Psi_0 \rangle} \approx f_{\mathcal{O}}(\vec{A}).
\label{generalnog}
\end{equation}
$\hat{\mathcal{O}} $ is an arbitrary operator,  $\{A_s \} = A_1, A_2, \ldots,  A_M$ $(\equiv \vec{A})$ are the relevant mean-fields, i.e. the expectation values of the 
corresponding single-particle operators $\{ \hat{A}_s \} = \hat{A}_1, \ldots, \hat{A}_M$, and both  $\hat{H}$ and $| \Psi_0 \rangle$ are usually $\vec{A}$- dependent. Consequently, $f_{\mathcal{O}}(\vec{A})$ is a $\mathbb{C}$-valued function of the mean-fields.   Each prescription of the form (\ref{generalnog}) will be termed the \textit{renormalization scheme} (RS).

Note, that usually 
the  RS (\ref{generalnog})  may be given a more specific form  
\begin{equation}
\langle  \hat{\mathcal{O}} \rangle_{G}  \equiv \frac{\langle \Psi | \hat{\mathcal{O}}| \Psi \rangle }{\langle \Psi | \Psi \rangle} \approx g^{\mathcal{O}}(\vec{A}) \frac{\langle \Psi_{0} | \hat{\mathcal{O}}| \Psi_{0} \rangle }{\langle \Psi_{0} | \Psi_{0} \rangle} \equiv g^{\mathcal{O}}(\vec{A})  \langle  \hat{\mathcal{O}} \rangle.
\label{g O general}
\end{equation}
In the above,  $g^{\mathcal{O}}(\vec{A})$ is termed   \textit{renormalization factor}.

There are many different renormalization schemes  (\ref{generalnog})   proposed in the literature. The  simplest take into account  only on the local (site-dependent) mean fields (e.g. local charge density or magnetization), \cite{Edegger}-\cite{Marcin Letters} 
The more advanced include also the inter-site  quantities (mean fields defined on bonds, e.g. pairing amplitude), cf. the Refs. \cite{Fukushima},  \cite{Sigrist}-\cite{Kai}. 
  In particular, the scheme of Ref. \cite{Sigrist} has been used in  Refs.  \cite{Marcin 1} and \cite{Marcin Polonica} and also in our previous work, \cite{Jedrak Spalek}. 
 
%
 
In this paper we focus our attention mainly on the renormalization scheme of. Ref. \cite{Fukushima}, which,  in our opinion, is the most promising one devised so far.
First,  due to the presence of fugacity factors  in  (\ref{Correlated state}), the Gutzwiller projection does not change the local densities of spin up (down) electrons, i.e. $\langle \hat{n}_{i\sigma} \rangle_{G} = \langle \hat{n}_{i\sigma} \rangle \equiv n_{i\sigma} $.  
Next, this formalism may, in principle, be  systematically  extended  beyond the second order in the inter-site  quantities  (formulas of Ref. \cite{Fukushima}  are provided up to this order). 
Also, it can be  relatively easily  applied to the extensions of Hamiltonian (\ref{t-J exact}), or to the  more complicated symmetry - breaking situations.
  

Suppose, that we have constructed the RMFT   specifying   appropriate single-particle mean-field  Hamiltonian $\hat{H}$. Next, we have to determine the optimal values of mean-fields appearing in a model. One  possibility is an application of the non-variational self-consistent approach based on the Bogoliubov-de Gennes (BdG) equations.
 However, this route, apart from its other drawbacks, \cite{JJJS_arx_0},   encounters a serious difficulty in the following sense.  Namely, the   renormalization scheme of that Ref.  \cite{Fukushima} is in general  not of the form  (\ref{g O general}). 
This feature does not allow for an unambiguous identification of the renormalization factors and  hence   the construction of the  effective MF Hamiltonian. Namely, there is no clear way to ascribe the corresponding single-particle operator expression  to the expectation values (\ref{generalnog}).  
Even if some way of replacing $ \langle  \hat{\mathcal{O}} \rangle$ (\ref{generalnog}) by the operator  counterpart may seem more natural than others, there is in fact no unique way of carrying out such procedure. Also,  different  such ways obviously  yield different versions of  RMFT Hamiltonians and consequently different BdG self-consistent equations, hence different predictions of the model. 
   
This  lack of uniqueness does not appear within  the variational method proposed by us recently \cite{JJJS_arx_0}, as  will be   discussed in detail below.  

The main aim of  the present paper is twofold. First, we  compare different  renormalization schemes  within the RMFT effective single-particle picture for the t-J model.  Moreover, as in  to Ref. \cite{Jedrak Spalek}, we compare   two distinct  methods of solving such MF models: the present  variational approach (labeled as \textit{var}) and a non-variational one, based on the Bogoliubov-de Gennes self-consistent equations  (labeled as \textit{s-c}).

The paper is organized as follows. In Section \ref{formalism} we  present those  parts  of our  approach   which are relevant to the present discussion.  In  Section \ref{results}  we present our numerical results, first for the d-wave superconducting (dRVB), (\ref{dRVB} and \ref{CHC}), and then  for the staggered-flux (SF) solutions (\ref{SF}), respectively.   In Appendix A  we provide   some of the technical details omitted in  main text. Section \ref{summary} contains a concluding remarks.

\section{Formalism: the method \label{formalism}}

\subsection{Self-consistent variational approach to the mean-field models \label{self cons var}}

The most of standard mean-field models  may be solved by employing two methods.  First, we may invoke the variational procedure, i.e. minimization of the MF grand potential or ground state energy, with respect to the values of the  order parameters (mean fields). Secondly, by using the self-consistency  (Bogoliubov-de Gennes) equations, expressing the basic fact that the mean-fields are averages of the corresponding operators. However,  those two routes are equivalent only  for   MF Hamiltonians of the   Hartree-Fock type (e.g. the BCS  Hamiltonian) \cite{JJJS_arx_0, FB}. 
In the case of RMFT for the t-J model,  the self-consistency of the MF formalism may be spoiled by unwary application of such  variational procedure, for both  nonzero temperatures and for $T=0$. This due to the non-Hartree character of the RMFT Hamiltonian, caused by MF treatment of the Gutzwiller projection.   

In such situation, the  non variational self-consistent method based on the Bogoliubov-de Gennes (BdG) equations is usually  applied to those advanced versions of RMFT, \cite{Didier, Sigrist, Marcin Letters, Marcin 1, Marcin Polonica}. However, this route  suffers from serious drawbacks \cite{JJJS_arx_0}. 

The other solution is the appropriate  modification of the variational method. Examples of  such self-consistent variational MF approach may be found  in  Refs. \cite{LiZhouWang, Liang Ren, Ogata Himeda, Kai}, 
 \cite{Wang F C Zhang}-\cite{Fukushima 2}.
The approaches  of those References  exhibit  various degree of generality,  and  differ  from each other with respect to technical details. Yet, all  are based on the variational principle of quantum mechanics (minimization of the expectation value of the Hamiltonian), applied to the MF case. As such, they are applicable to $T=0$ situation only. 

Our point of departure is different. We base our approach  (\cite{Jedrak Spalek, JJJS_arx_0}) on the   \textit{maximum entropy principle} (MaxEnt) \cite{Jaynes}, which  is   the basis of Bayesian mathematical statistics. In particular, it may be viewed  also as the basis of the standard (non mean-field) statistical mechanics \cite{Jaynes, statistical mechanics}. The application of MaxEnt inference  to the non-standard case of MF  statistical-mechanical description is a  natural   extension  of this fundamental principle (\cite{JJJS_arx_0}, c.f. also \cite{Argentynczycy}). To  ensure the  self consistency of MF description, it is necessary to introduce  additional constraints, not present in the standard statistical mechanics. This could be achieved in the most natural manner by the Lagrange multiplier method. Explicitly,  starting from an arbitrary MF Hamiltonian $\hat{H}(\vec{A})$, we define yet another  MF Hamilton operator $\hat{H}_{\lambda}$ according to\footnote{Lagrange multipliers $\lambda_s$ appearing in (\ref{GMF Hamiltonian with lambdas}) should not be confused with the fugacity factors $\lambda_{i \sigma}$   appearing in Eqn. (\ref{Correlated state})}
\begin{equation}
\hat{H} ~~\to ~~ \hat{H}_{\lambda} ~= ~  \hat{H}-\sum_{s=1}^{M}   \lambda_{s}(\hat{A}_{s} - A_{s} ).
\label{GMF Hamiltonian with lambdas}
\end{equation}
The correct grand-canonical MF density operator is then given by
\begin{equation} 
 \hat{\rho}_{\lambda} =  \mathcal{Z}_{\lambda}^{-1} \exp \big(-\beta ( \hat{H}_{\lambda} - \mu  \hat{N})\big), ~~~  \mathcal{Z}_{\lambda}= \text{Tr}[\exp \big(-\beta ( \hat{H}_{\lambda} - \mu  \hat{N})\big)].
\label{correct MF density matrix}
\end{equation}
Next,  the generalized grand-potential Landau functional  is defined as
\begin{eqnarray}
\mathcal{F}(\vec{A}, \vec{\lambda}) &\equiv&   -\beta^{-1} \ln( \text{Tr} [e^{-\beta(\hat{H}_{\lambda} -\mu \hat{N}) }]) =  -\beta^{-1} \ln(\mathcal{Z}_{\lambda}),
\label{mathcalF tJ BCS like}
\end{eqnarray}
with the inverse temperature $\beta = 1/k_{B}T$. 
The equilibrium values of $\vec{A}= \vec{A}_0$, $\vec{\lambda} = \vec{\lambda}_0$ are  the solution of the set of  equations
\begin{equation}
\nabla_{A} \mathcal{F} =0, ~~ ~~ \nabla_{\lambda} \mathcal{F} =0,
\label{derivative of mathcalF A, lambda}  
\end{equation} 
for which $\mathcal{F}(\vec{A}, \vec{\lambda}) $ has the lowest value. 
By taking the derivatives  with respect to $\vec{\lambda}$ only, and subsequently putting $\vec{\lambda}= \vec{0}$,
\begin{equation}
 \nabla_{\lambda} \mathcal{F} =0,  ~~~~~\vec{\lambda}= \vec{0},
\label{derivative of mathcalF lambda}  
\end{equation}
we obtain standard Bogoliubov-de Genes (BdG) self-consistent equations. In such a case we denote the chemical potential $\tilde{\mu}= \mu_{s-c}$, which corresponds to $\tilde{\mu}= \mu + \lambda$ for the \textit{var} method. The thermodynamical grand potential $\Omega$ and the free energy $F$ are defined respectively as 
\begin{equation}
\Omega(T, V, \mu) = \mathcal{F}(T, V, \mu; \vec{A}_0(T, V, \mu),  \vec{\lambda}_0(T, V, \mu)  ), ~~~~~F = \Omega + \mu N 
\label{Thermo potentials}
\end{equation}
for \textit{var}, and similarly (with $\mu_{s-c}$, $ \vec{\lambda}_0  =  \vec{0}$) for \textit{s-c} method.

Obviously, the present formalism is valid for any  non-zero temperature, but not  for   $T=0$.  Consequently,  we should replace  pure states $|\psi_0 \rangle $ in Eqs. (\ref{Correlated state}) -  (\ref{g O general}) by mixed ones (\ref{correct MF density matrix}), $  \hat{\rho}_{\lambda}$.  

On the other hand, the Gutzwiller approximation in a form (\ref{generalnog}) or (\ref{g O general}) is devised for a ground  state, hence for temperature $T=0$. However, the solutions of the mean field models (and the RMFT of the  t-J model in particular), obtained for non-zero,  but sufficiently low $T$ are   practically identical to those of  the real $T=0$ analysis. Consequently, $  \hat{\rho}_{\lambda}$ is practically indistinguishable from $|\psi_0 \rangle $,  and thus the application of the finite-temperature formalism in the $\beta \to \infty$ limit is fully justified.

Reader interested in details of our method may  consult  Ref. \cite{JJJS_arx_0}. Below we recall some of its features, which are important from the point of view of the present application. They are justified in the Appendix A. 

First,  the $\vec{A}$ - dependence  of the nontrivial\footnote{i.e. not proportional to the unit operator} part of the  Hamiltonian $\hat{H}_{\lambda}(\vec{A})$ (\ref{GMF Hamiltonian with lambdas}) is completely determined by the  set   of the  single-particle operators out of which   $\hat{H}_{\lambda}(\vec{A})$ and $\hat{H}(\vec{A})$ are  composed.  
  
Moreover,  two  MF  Hamiltonians $\hat{H}^{(1)}$ and $\hat{H}^{(2)}$ (or $\hat{H}_{\lambda}^{(1)}$ and $\hat{H}_{\lambda}^{(2)}$),  having the same nontrivial  operator  part and the same  $\vec{A}$ -dependence of the expectation value, i.e. $\langle \hat{H}^{(1)}_{\lambda} \rangle = \langle \hat{H}^{(1)}  \rangle = \langle \hat{H}^{(2)}_{\lambda} \rangle = \langle \hat{H}^{(2)}  \rangle$,  are   equivalent. Namely, they yield identical  equilibrium values of mean fields, quasi-particle energies, thermodynamic potentials etc. 
This feature guarantees that, when constructing RMFT Hamiltonian,   different assignments of the   operator  expression to the expectation values  (\ref{aver hopping Fuk}) lead to identical results.  
 
Parenthetically, the values of the Lagrange multipliers $\vec{\lambda}$ may differ between two such equivalent MF Hamiltonians $\hat{H}^{(1)}_{\lambda}$, $\hat{H}^{(2)}_{\lambda}$. This indicates, that   $\vec{\lambda}$ have no unambiguous physical interpretation  by themselves, but only in the certain combinations with the mean fields.  Only the Lagrange multipliers related to the quantities of \textit{a priori}  known average values, e.g. particle number  $N$,  are identical for all such equivalent Hamiltonians.   

\subsection{Application to  the  renormalized mean-field t-J model \label{application to RMFT}}

For the RS that can be given the form (\ref{g O general}), with $\lambda_{i \sigma} = 1$ in (\ref{Correlated state}),  and in the absence of the long-range antiferromagnetic order, the mean-field Hamiltonian $\hat{H} $  may be taken in a form \cite{Didier, Marcin Letters, Sigrist, Marcin 1, Marcin Polonica} 
\begin{eqnarray}
\hat{H} &=&  \sum_{\langle i j \rangle \sigma}\big (t_{ij} g^{t}_{ij} c_{i \sigma}^{\dag} c_{j \sigma} +  \text{H.c.} \big) - \frac{3}{4} J_{ij} g^{J}_{ij} (\chi_{ji}  c_{i \sigma}^{\dag} c_{j \sigma} + \text{H.c.} - |\chi_{ij}|^{2})  \nonumber \\   &-& \sum_{\langle i j \rangle \sigma} \frac{3}{4} J_{ij} g^{J}_{ij} (\Delta_{ij}   c^{\dag}_{j \sigma}c_{i -\sigma}^{\dag} + \text{H.c.}   - |\Delta_{ij}|^{2}).
\label{ren tJ Ham}
\end{eqnarray}
In the above expression,  $c_{i \sigma}^{\dag} $ ($ c_{j \sigma}$) are ordinary fermion creation (annihilation) operators, $\chi_{ij} = \langle  c_{i \sigma}^{\dag}  c_{j \sigma} \rangle$, and $\Delta_{ij} = \langle  c_{i -\sigma}  c_{j \sigma} \rangle = \langle  c_{j -\sigma}  c_{i \sigma} \rangle$ are respectively, the hopping amplitude (bond-parameter) and the RVB gap parameter, both taken for nearest neighbors $\langle i j \rangle $.  $\langle \hat{\mathcal{O}}\rangle$ denotes the average value of the operator $\hat{\mathcal{O}}$  evaluated with the help of a MF state\footnote{This corresponds to $\langle \hat{\mathcal{O}}\rangle_0$ in notation of Ref. \cite{Fukushima}} (\ref{Correlated state}). 

In (\ref{ren tJ Ham}), the interaction term ($ \mathbf{S}_{i}\cdot \mathbf{S}_{j} $) of Hamiltonian (\ref{t-J exact}) has been treated within  the  Hartree - Fock decoupling in both the particle - particle ($\Delta_{ij}$) and particle - hole ($\chi_{ij}$) channels.  The renormalization factors $g^{t}_{ij} $ and $g^{J}_{ij} $,  
are given by (\ref{g O general}), with   $\hat{\mathcal{O}} = \sum_{\langle i j \rangle \sigma} (g^{t}_{ij} c_{i \sigma}^{\dag} c_{j \sigma} +  \text{H.c.})$ and $\hat{\mathcal{O}} =  \mathbf{S}_{i}\cdot \mathbf{S}_{j} $, respectively. Their explicit form depends on the approximation used to obtain r.h.s. of (\ref{g O general}).

As mentioned above, the non-Hartree-Fock   character of RMFT Hamiltonians $\hat{H}$ is due to MF treatment of the Gutzwiller projection. Consequently, in  order to obtain $\hat{H}_{\lambda}$,  we must add to  $\hat{H} $ the constraints corresponding to the   mean fields appearing in $g^t$ and $g^J$.  

The simplest   renormalization factors \cite{Edegger}-\cite{Marcin Letters}  depend  solely on local hole densities  $x_i = 1- n_i$, i.e.
\begin{equation}
 g^{t}_{ij} = \sqrt{\frac{4x_i x_j}{(x_i + 1)(x_j + 1)}} ~~~~\text{and} ~~~ g^{J}_{ij} = \frac{4}{(x_i + 1)(x_j + 1)}.
\label{g ren fac x only}
\end{equation} 
In the framework of our method, in the   homogeneous case $x = x_i $,  the  additional Lagrange multiplier coupled to the total particle number  is introduced.  For a non-homogeneous case, local chemical potentials should be introduced, which may be found in Refs.  \cite{LiZhouWang, Liang Ren, Fukushima 2}. 

In the  case of more complicated form of $g^{t}$, $g^{J}$ factors, depending apart form $x_i$ also on the inter-site mean-field variables (\cite{Fukushima, Sigrist, Marcin 1, Marcin Polonica}), the MF Hamiltonian has to be modified further. The Lagrange multipliers related   to the average hopping (bond order) and superconducting order parameters acquire non-zero values in the equilibrium situation.

 As pointed out previously, the form of Gutzwiller approximation of Ref.  \cite{Fukushima} does not reduce to multiplication by renormalization factors, (\ref{g O general}). E.g., for the hopping amplitude,   we have  
\begin{equation}
 \langle c_{i \uparrow}^{\dag} c_{j \uparrow}  \rangle_{G}   \approx  \sqrt{\frac{1-n_i}{1 - n_{i \uparrow}}}  \sqrt{\frac{1-n_j}{1 - n_{j \uparrow}}}  \left( \chi_{ij \uparrow}  -  \chi_{ij \downarrow}  \frac{\chi_{ij \uparrow} \chi^{\ast}_{ij \downarrow} + \Delta_{ji} \Delta^{\ast}_{ij} }{(1 - n_{i\downarrow})(1 - n_{j\downarrow})} \right),
\label{aver hopping Fuk}
\end{equation}
(Eqn. (15) of Ref. \cite{Fukushima}, but with  different  notation). This,  in general,  does not allow for    identification of renormalization factor $g^{t}$.
 However, for completeness, we want to  compare the variational approach with the  non-variational treatment of RMFT based on Ref. \cite{Fukushima}. Hence, using (\ref{aver hopping Fuk}) we  may quite reasonably define those factors for a simple homogeneous  non-magnetic states ($ n_{i \uparrow} = n_{i \downarrow} = n_{i}/2$, $ \langle  c_{i \uparrow}^{\dag}  c_{j \uparrow} \rangle = \chi_{ij \uparrow} = \chi_{ij \downarrow} \equiv \chi_{ij}$) as
\begin{equation}
(I)~~~~~~~~ g^{t}_{ij} =   
    \sqrt{\frac{1-n_i}{1 - \frac{n_{i}}{2}}}  \sqrt{\frac{1-n_j}{1 - \frac{n_{j}}{2}}}  \left( 1  -   \frac{\chi_{ij } \chi^{\ast}_{ij  } + \Delta_{ji  } \Delta^{\ast}_{ij} }{(1 - \frac{n_{i}}{2})(1 - \frac{n_{j}}{2})} \right), 
\label{gt ren fac x chi delta Fuk}
\end{equation}
\begin{equation}
g^{J}_{ij} =  \frac{1}{(1- \frac{n_{i}}{2}) (1- \frac{n_{j}}{2})}.
\label{gJ ren fac x chi delta Fuk}
\end{equation} 
The prescription (\ref{gJ ren fac x chi delta Fuk}) will be referred to as a renormalization scheme (I). 
This will  be  confronted with   $g$-factors taken from Ref. \cite{Sigrist}
\begin{equation}
(II)~~~ ~~~~~~~ g^{t}_{ij} = \sqrt{\frac{4x_i x_j(1-x_i)(1-x_j)}{(1-x_i^{2})(1-x_j^{2}) + 8(1-x_i x_j)|\chi_{ij}|^{2} + 16|\chi_{ij}|^{4}}}, 
\label{gt ren fac x chi delta}
\end{equation}
\begin{equation}
 g^{J}_{ij} = \frac{4(1-x_i)(1-x_j)  }{(1-x_i^{2})(1-x_j^{2}) + 8x_i x_j(|\Delta_{ij}|^{2}-|\chi_{ij}|^{2}) + 16(|\Delta_{ij}|^{4}+|\chi_{ij}|^{4})},
\label{gJ ren fac x chi delta}
\end{equation}
 referred to as renormalization scheme (II). 
Usually within the RMFT,  the physical (renormalized) superconducting parameter is defined as $g^{t}\Delta$ instead of bare $\Delta$ itself, \cite{Edegger}. This is also the case for scheme (II).
On the other hand, within the scheme (I), there appear a  separate expression for renormalized  value of $ \langle  c_{i -\sigma}  c_{j \sigma} \rangle_G = \langle  c_{j -\sigma}  c_{i \sigma} \rangle_G$, cf. Eqn. (18) of \cite{Fukushima}. Again, for simple homogeneous, non-magnetic states, the  corresponding $g^{\Delta}$ factor can be quite reasonably identified as
\begin{equation}
(I)~~~~~~~~ g^{\Delta}_{ij} =   
    \sqrt{\frac{1-n_i}{1 - \frac{n_{i}}{2}}}  \sqrt{\frac{1-n_j}{1 - \frac{n_{j}}{2}}}  \left( 1 +  \frac{\chi_{ij } \chi^{\ast}_{ij  } + \Delta_{ji} \Delta^{\ast}_{ij} }{(1 - \frac{n_{i}}{2})(1 - \frac{n_{j}}{2})} \right).  
\label{gDelta ren fac x chi delta Fuk}
\end{equation}
Note, that (\ref{gDelta ren fac x chi delta Fuk}) differs from (\ref{gt ren fac x chi delta Fuk}). 



\section{Results: d-wave superconducting and staggered flux phase \label{results}}
Below we present our numerical  results, first for the plain d-wave superconducting state with no magnetic order (dRVB), and next for the staggered-flux (SF) normal solution. Even for those  simplest MF states, the results obtained within various renormalization schemes and/or the solving methods  differ remarkably.

\subsection{d-wave superconducting (dRVB) solution at $x = \frac{1}{8}$  \label{dRVB}}
 The solution analyzed here\footnote{Part of the results presented in this section may be found also in \cite{Jedrak Spalek}.} is constructed to possess   full symmetry of the underlying square lattice,  only the  superconducting order parameter is  assumed to have   $d_{x^2 - y^2}$ symmetry. 
Consequently,  we are left with three independent mean fields $\vec{A} = (n, \chi,   \Delta )$, where  $  \chi_x = \chi = \chi_y$, $\Delta_x = \Delta = - \Delta_y$, and the same number of the corresponding Lagrange multipliers, $\vec{\lambda} = (\lambda, \lambda^{\chi}, \lambda^{\Delta})$, where $  \lambda^{\chi} = \lambda^{\chi}_x = \lambda^{\chi}_y$, $ \lambda^{\Delta} = \lambda^{\Delta}_x  = -  \lambda^{\Delta}_y$, and  $n = n_i = \sum_{\sigma} \langle  c_{i \sigma}^{\dag}  c_{i \sigma} \rangle $,   $\chi_{ij} = \chi_{\tau}$, $\tau = x (y)$ for the bonds in $x$ ($y$) directions, respectively. Also,  $\sqrt{2} \Delta_{ij}= \Delta_{\tau} $, and all the above quantities are taken as real. 
Diagonalization of (\ref{GMF Hamiltonian with lambdas})  in the present case yields
\begin{equation} 
\hat{H}_{\lambda} - \mu \hat{N} =  \sum_{\mathbf{k}} E_{\mathbf{k}} (\hat{\gamma}^{\dag}_{\mathbf{k}0}\hat{\gamma}_{\mathbf{k}0} + \hat{\gamma}^{\dag}_{\mathbf{k}1}\hat{\gamma}_{\mathbf{k}1}) + \sum_{\mathbf{k}} (\xi_{\mathbf{k}} - E_{\mathbf{k}}) + C,
\label{MF tJ BCS like}
\end{equation}
with 
\begin{equation}
E_{\mathbf{k}} = \sqrt{\xi^2_{\mathbf{k}} + D^2_{\mathbf{k}}},
\label{E_k}
\end{equation} 

\begin{equation}
D_{\mathbf{k}} = \sqrt{2}\sum_{\tau}D_{\tau}\cos(k_{\tau}), ~~~~~ \xi_{\mathbf{k}} = -2\sum_{\tau}T_{\tau}\cos(k_{\tau})  -\mu - \lambda.
\label{E_k}
\end{equation} 
Also,
\begin{equation} 
T_{\tau} = -t_{1\tau} g^{t}_{1\tau} + \frac{3}{4}J_{\tau} g^{J}_{\tau}\chi_{\tau} + \lambda^{\chi}_{\tau}, ~~~~D_{\tau} =  \frac{3}{4}J_{\tau} g^{J}_{\tau}\Delta_{\tau} + \lambda^{\Delta}_{\tau}, 
\label{T tau}
\end{equation} 
\begin{equation} 
\frac{C}{ \Lambda } =\lambda n + \sum_{\tau} \big( \frac{3}{4} J_{\tau} g^{J}_{\tau} (2 \chi^{2}_{\tau} + \Delta^{2}_{\tau}) + 4 \chi_{\tau} \lambda^{\chi}_{\tau} + 2\Delta_{\tau}\lambda^{\Delta}_{\tau}\big).
\label{T tau C}
\end{equation}
The Gutzwiller renormalization factors read now, respectively
\begin{equation} 
(I)~~  ~g^{t}_{\tau}(n, \chi_{\tau},  
\Delta_{\tau}) = \frac{2 (1-n)}{2-n} \left(1-\frac{4 \chi_{\tau}^2 + 2 \Delta_{\tau}^2}{(2-n)^2}\right), ~~~~ g^{J}_{\tau}(n) =  \frac{4}{(2-n)^2},  \nonumber
\label{g t tau old}
\end{equation}
 \begin{equation} 
  g^{\Delta}_{\tau}(n, \chi_{\tau},  
\Delta_{\tau}) = \frac{2 (1-n)}{2-n} \left(1+\frac{4 \chi_{\tau}^2 + 2 \Delta_{\tau}^2}{(2-n)^2}\right), 
\label{g t tau}
\end{equation}

\begin{equation} 
(II)~~~~~~~~~~~~ g^{t}_{\tau}(n, \chi_{\tau}) = \frac{2n(1-n)}{n(2-n) + 4 \chi^{2}_{\tau}}, \nonumber
\label{g t tau s}
\end{equation}
\begin{equation} 
g^{J}_{\tau}(n, \chi_{\tau}, \Delta_{\tau}) =\frac{4n^2}{n^2(2-n)^2 + (1-n)^2 (4\Delta^2_{\tau} - 8 \chi^{2}_{\tau}) + 4\Delta^4_{\tau} + 16\chi^4_{\tau}}.
\label{g J tau s}
\end{equation}
Note, that for (II) $ g^{\Delta}_{\tau} =  g^{t}_{\tau}$. The generalized Landau functional is given by
\begin{eqnarray}
\mathcal{F}(\vec{A}, \vec{\lambda}) & = &C(\vec{A}, \vec{\lambda}) + \sum_{\mathbf{k}}\big(  (\xi_{\mathbf{k}} - E_{\mathbf{k}}) - \frac{2}{\beta} \ln\big(1 + e^{-\beta E_{\mathbf{k}}}\big)\big), 
\label{mathcalF tJ BCS like BCS}  
\end{eqnarray}
from which the explicit form of Eqs. (\ref{derivative of mathcalF A, lambda}) can be easily obtained, but are not presented here  in an explicit form .

\textit{Numerical results.} We begin with the  analysis of dRVB solution for the 'magic doping', $x = \frac{1}{8} = 0.125$. The equations (\ref{derivative of mathcalF A, lambda}) and (\ref{derivative of mathcalF lambda}) are solved for the lattice of $\Lambda = \Lambda_x \Lambda_y$ sites, $\Lambda_x = \Lambda_y = 256$, with   $ J_{\tau} = 1$ $t_{\tau} = -3 J$  and for low temperature, $\beta = 500$. Both   dRVB solution, as well as isotropic normal state (Fermi sea, (FS), not discussed explicitly here) are present. The staggered flux state, expected for lower doping, have not been found (\textit{var}) or is not stable against FS (\textit{s-c})  for  $x =  0.125$.  

In Tables I. and II. we give the values of the thermodynamic potentials,  mean-fields and molecular fields for  dRVB solutions obtained within both methods (\textit{var}, \textit{s-c}) and both renormalization schemes ((I), (II)).
\begin{center} 
Table I. Values of the thermodynamic potentials (per site) for dRVB solutions. $\tilde{\Omega}$ ($F$) stands for  $  \Omega -\lambda N$ ($\Omega + \mu N$)  for  \textit{var} and $\Omega_{s-c}$ ($\Omega_{s-c} +\mu_{s-c} N$) for \textit{s-c} methods, respectively.\\ 
\begin{tabular}{c c c c c}\hline \hline
Therm. Pot. & \textit{var} (I)  &  \textit{var} (II)   & \textit{s-c} (I)  & \textit{s-c}  (II)        \\  \hline 
$\Omega/\Lambda $  & -6.0444393   & -5.7586779   &  -  &  - \\  
$\tilde{\Omega}/\Lambda$   & -1.0897421  &  -1.0766359 &  -1.01117344   & -1.03614582  \\  
 $F/\Lambda $   &   -1.343195431  &   -1.366146003 &  -1.339864247  &  -1.364716747   \\  
\hline \hline
\end{tabular}
\end{center}
 
For a fixed particle concentration  $n$,   the relevant thermodynamical potential  is  the free energy $F$. By construction of the solution the   value  of $F$  obtained within each RS is alway lower for the \textit{var} method then for the  \textit{s-c} one. The free energy  is also the quantity, that determines which   solution of (\ref{derivative of mathcalF A, lambda}) or  (\ref{derivative of mathcalF lambda})   corresponds to the stable equilibrium situation.  However,  by no means it may be used to   favor one or another renormalization scheme. For example, if we compare the values of the $F$ for \textit{var}  method in the present case, and also invoke its value $F^{0}_{var}$, corresponding to the simplest Gutzwiller factors  (given by  Eqn. (\ref{g ren fac x only}), and not analyzed explicitly here, c.f. however Ref. \cite{Marcin Polonica}), we see that $F^{0}_{var} = -1.5070 < F^{II}_{var} < F^{I}_{var}$. Clearly, it does not mean that RS defined by (\ref{g ren fac x only})  should be preferred over   (I) or (II). 

\begin{center}
Table II. Values of the equilibrium chemical potentials and MF parameters ($ \vec{A}_0$, $ \vec{\lambda}_0$) for dRVB solutions.
$\tilde{\mu} $ stands for $\lambda + \mu $ (\textit{var}), and for $\mu_{s-c}$ (\textit{s-c}).\\
\begin{tabular}{c c c c c}\hline \hline

Variable & \textit{var} (I)  &  \textit{var} (II)   & \textit{s-c}  (I)  & \textit{s-c}  (II)   \\  \hline 
$\mu$  & 5.37285   &  5.02004  &  -    &  -   \\  
$\lambda $   &  -5.66251    & -5.35091 &  -   & -  \\  
$\tilde{\mu}$   & -0.28966    &  -0.33087   &   -0.37565    &  -0.37551    \\   
$\chi_x $   &   0.19414   &  0.18807     &     0.19011   &    0.19074     \\  
$\lambda^{\chi}_{x} $   &  -0.15883    & -0.16985    & -   &  -\\
$\frac{1}{\sqrt{2}} \Delta_x  $   &   0.10897   & 0.13200    &  0.12565   &  0.12345  \\ 
 $\frac{1}{\sqrt{2}}\lambda^{\Delta}_{x} $   &  -0.08915     &  -0.01111   &  - &  -\\ 
$2T_{x} $   & 1.72718       &  1.69732     &  2.01576       &   2.04556  \\  
$\sqrt{2}D_{x}$   &   0.33830    &   0.58926    &    0.59565    &   0.57209  \\   
\hline \hline
\end{tabular}
\end{center}
 
Let us now analyze Table II. First,  within RS (II),    $|\lambda^{\chi}_{\tau}|$   is significantly larger than   $|\lambda^{\Delta}_{\tau}|$. This indicates that the Hamiltonian (\ref{MF tJ BCS like}) deviates from the Hartree-Fock form more  with respect to $\chi_{\tau}$ then with respect  to $\Delta_{\tau}$. This is   due to the particular form of $\chi_{\tau}$-dependent, $\Delta_{\tau}$-independent  renormalized hopping term. Similar conclusions are valid for RS (I), but then  the $\chi_{\tau}$- and $\Delta_{\tau}$ dependence of $\langle \hat{H} \rangle$ is more symmetric (c.f. Eqs. \ref{gt ren fac x chi delta Fuk} and \ref{gJ ren fac x chi delta Fuk}, or \ref{g t tau old}) and thus the difference between  $|\lambda^{\chi}_{\tau}|$ and $|\lambda^{\Delta}_{\tau}|$ is smaller then for RS (II).  

\begin{figure}[h]
\begin{center}
\rotatebox{270}{\scalebox{0.40}{\includegraphics{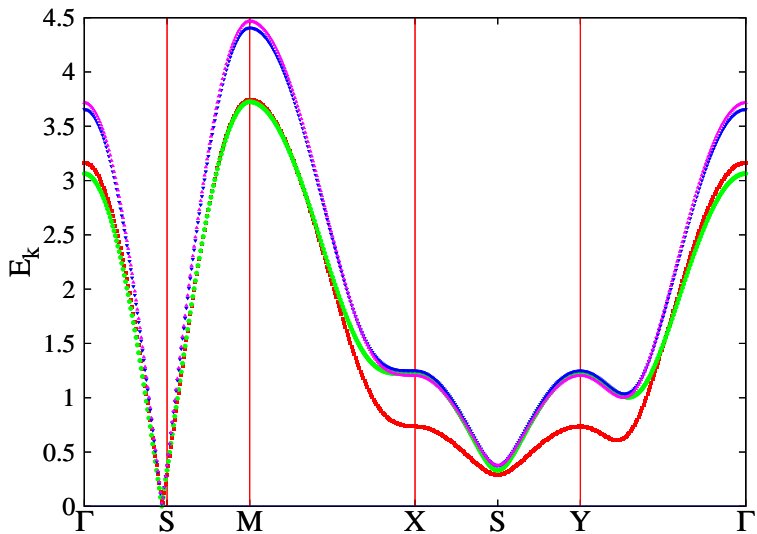}}}
\end{center}
\caption{(Color online) Dispersion relations along the main symmetry lines in the Brillouin zone for the dRVB solutions for a square lattice, of the size $\Lambda_x = \Lambda_y  = 256$, and for the filling $n=0.875$.   Triangles -  self-consistent, non-variational  results for (I) and (II), red squares (I) and green circles (II)- the present variational method.}
\label{dispertion for 256 RVB}
\end{figure}

Large values of $|\lambda^{\chi}_{\tau}|$ significantly affect (through $T_{\tau}$ and $D_{\tau}$, Eqn. (\ref{T tau}))   the quasi-particle spectra, $E_{\mathbf{k}}$, cf. Fig \ref{dispertion for 256 RVB}.   Namely, the excitation energies within our method are always \textit{lower} then those of BdG self-consistent  approach. This is due to relative minus sign between equilibrium values of $ \lambda^{\chi}_{\tau} $ and $\chi_{\tau}$  ($ \lambda^{\Delta}_{\tau} $ and $\Delta_{\tau}$),  cf.  also  Ref. \cite{Jedrak Spalek}. 
Although for RS (II) the differences between the methods (\textit{var}, \textit{s-c}) are pronounced mainly in the regions of the Brillouin zone which are far from the Fermi surface, the difference of the tangent at the cone near point $S$   may be of some significance. However, what is more important, RS (I) \textit{var} gives different excitation energies then other cases also  along the $X$-$Y$ direction, i.e. close to the Fermi surface ($\xi_{\mathbf{k}} = 0$).  The reason for such behavior is obvious from the analysis of Tab. II;   along $X$-$Y$ direction  $\xi_{\mathbf{k}} = \text{const}$ and the main contribution to $E_{\mathbf{k}}$ comes from $D_{\mathbf{k}}$, determined in turn by   $D_{\tau}$, which is exceptionally low for (I) \textit{var}. 

\subsection{Doping dependence of the mean-field quantities: critical hole concentrations  \label{CHC}}

\begin{figure}[h!]
\begin{center}
\rotatebox{270}{\scalebox{0.35}{\includegraphics{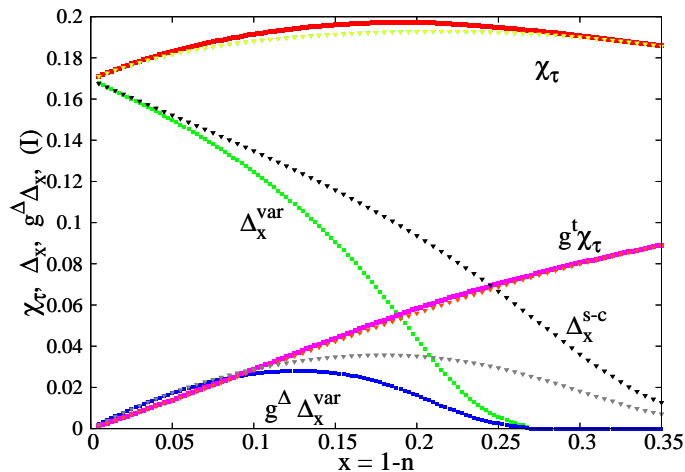}}}
\end{center}
\caption{(Color online). Renormalization scheme (I):  Doping dependence of the bond-order parameters $\chi_x=\chi_y$,  the superconducting order parameters $\Delta_x = - \Delta_y$, and their renormalized correspondents $g^{t}\chi_x=g^{t}\chi_y$ and $g^{\Delta}\Delta_x = - g^{\Delta}\Delta_y$,   both for the \textit{s-c} (triangles) and the \textit{var} (squares) methods.}
\label{x dependen Fuk}
\end{figure}

\begin{figure}[h!]
\begin{center}
\rotatebox{270}{\scalebox{0.35}{\includegraphics{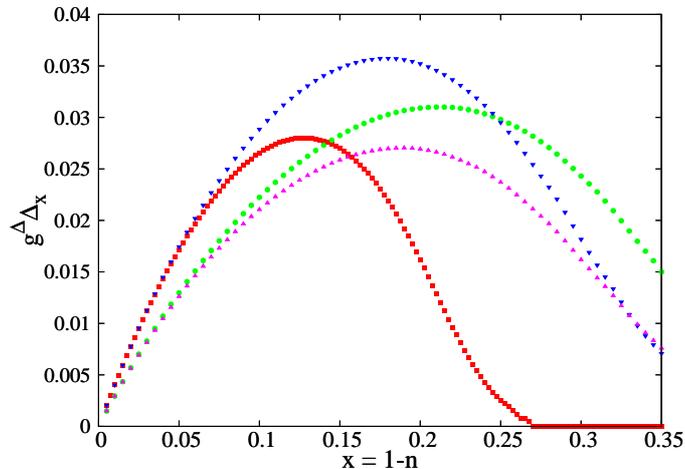}}}
\end{center}
\caption{(Color online). Doping dependence of renormalized superconducting order parameter  $g^{\Delta}\Delta_x$ for the paired (dRVB) state within various schemes described in the text. For $x=0.2$, from the top to the bottom: (I) \textit{s-c} (blue triangles), (II) \textit{var}  (green circles), (II) \textit{s-c} (violet triangles),  (I) \textit{var}  (red squares). Only the latter choice yields the (right value of the) critical doping  $x_c \approx 0.27$, above which the gap disappears.}
\label{doping dependance ren Deltas}
\end{figure}

Next, we are going to discuss the  changes appearing as the function of doping. Those are the most interesting results obtained in the present paper.

 The $(1-n)$-dependences of mean-fields $\chi_{\tau}$ and $\Delta_{x}$, as well as the physical (renormalized) gap parameter $g^{\Delta}\Delta_{x}$ and renormalized hopping $g^{t} \chi_{\tau}$ for scheme (I) are analyzed in Fig. \ref{x dependen Fuk}. (The analogous picture for RS (II)  is given in \cite{Jedrak Spalek}).
 Note, that for (I) \textit{var},    $\Delta_{x}$ vanishes at the  critical concentration $x_c \approx 0.27$. This  is in much better agreement with  the experimental results then the predictions of the other cases, for which $x_c > 0.35$. The situation  is illustrated   explicitly in Fig. \ref{doping dependance ren Deltas}, where the $x$-dependence of the $g^{\Delta} \Delta_{x}$ is shown  also for the scheme (II), for both methods.
 Also, the right  value of the  optimal hole concentration  ($x \approx 0.125$) is obtained for (I) \textit{var}, in contrast to either non-variational treatment or to RS (II).

\subsection{Staggered flux solution at $x \approx  \frac{1}{8}$ \label{SF}}
As a next example we analyze the staggered-flux  (SF) phase. 
This MF state   has a long history, first being proposed by Affleck and Marston \cite{Affleck Marston}, as a variational trial MF state for the Heisenberg model. It was intensively  investigated later, due to  its possible connection with the pseudo-gap state in cuprates, cf. e.g. \cite{Zhang SF}.

The SF  differs from a normal Fermi sea  (FS) solution by  the presence of complex hopping amplitude  $\chi_{ij} = |\chi|\exp( (-1)^{(\mathbf{i_x} + \mathbf{j_x})} i \varphi) \equiv \xi_1 \pm i \xi_2 $. Such $\chi_{ij}$ implies   existence of  circulating  currents,  which direction changes from plaquette to plaquette in an alternating fashion (orbital antiferromagnet), \cite{Zhang SF}. Consequently,  a two-sublattice structure emerges, with the unit cell of the size $\sqrt{2} a \times \sqrt{2} a $ in direct space and new (folded) Brillouin zone (NBZ).

Within the framework of our method we add appropriate constraints. This introduces, apart from $\lambda$ ascribed to $n$, also the complex Lagrange multiplier $\eta_{ij}   = \eta_1 \mp i \eta_2 $,  tailored to $\chi_{ij}$  (our sign  convention for $\eta_{ij} $ for each bond is opposite to that  for $\chi_{ij}$). Thus,  we have three independent real mean fields $\vec{A} = (n, \xi_1, \xi_2)$, and   the same number of the corresponding real Lagrange multipliers, $\vec{\lambda} = (\lambda, \eta_1, \eta_2)$. 
In the present case the renormalization factors (\ref{gJ ren fac x chi delta Fuk}) and (\ref{gJ ren fac x chi delta}) read, respectively
\begin{equation}
(I)~~~~ g^{t}(n, \xi_1, \xi_2) = \frac{2 (1-n)}{2-n} \left(1-\frac{4 \left(\xi _1^2+\xi_2^2\right)}{(2-n)^2}\right), ~~~ g^{J}(n) = \frac{4}{(2-n)^2}.
\end{equation}
\begin{equation}
(II)~~~~~~~~~~~~~ g^{t}(n, \xi_1, \xi_2) = \frac{2n(1-n)}{n(2-n) + 4 (\xi^2_1 + \xi^2_2)}, \nonumber
\end{equation}
\begin{equation}
g^{J}(n, \xi_1, \xi_2) = \frac{4n^2}{n^2(2-n)^2  - 8(1-n)^2 (\xi^2_1 + \xi^2_2) + 16 (\xi^2_1 + \xi^2_2)^2}.
\end{equation}
Instead of Eqs. (\ref{MF tJ BCS like}) - (\ref{mathcalF tJ BCS like}) we have now
\begin{equation} 
\hat{H}_{\lambda} - \mu \hat{N} = C(\vec{A}, \vec{\lambda}) +  \sum^{NBZ}_{\mathbf{k} s \sigma} E_{\mathbf{k} s \sigma} \hat{\alpha}^{\dag}_{\mathbf{k} s \sigma } \hat{\alpha}_{\mathbf{k}s \sigma },
\label{MF tJ BCS like SF}
\end{equation}
where $E_{\mathbf{k} s \sigma} = -\tilde{\mu} + s \sqrt{\epsilon^2_{\mathbf{k}} + \chi^2_{\mathbf{k}}}$, $\epsilon_{\mathbf{k}} =  T_{1}\gamma_{+}(\mathbf{k})$, and 
$\chi_{\mathbf{k}} = T_{2}\gamma_{-}(\mathbf{k})$.  Also, $\gamma_{\pm}(\mathbf{k}) \equiv 2(\cos(k_x) \pm \cos(k_y) )$  and
\begin{equation} 
T_{1} =  t g^{t}  + \frac{3}{4}J  g^{J} \xi_1 + \eta_1, ~~~~T_{2} =  \frac{3}{4}J  g^{J} \xi_2 + \eta_2,
\label{T tau 1 SF}
\end{equation} 
\begin{equation} 
C = \Lambda \left( \lambda n + 3 g^{J} (\xi^2_1 + \xi^2_2) + 8 (\xi_1 \eta_1 + \xi_2 \eta_2) \right).
\label{T tau SF}
\end{equation}
The generalized Landau functional (\ref{mathcalF tJ BCS like}) takes now the following form
\begin{eqnarray}
\mathcal{F}(\vec{A}, \vec{\lambda}) & =&  C(\vec{A}, \vec{\lambda}) - \frac{1}{\beta} \sum^{NBZ}_{\mathbf{k} s \sigma }  \ln\big(1 + e^{-\beta E_{\mathbf{k}  s \sigma}}\big).
\label{mathcalF tJ BCS like SF}
\end{eqnarray}

\textit{Numerical results}. As mentioned in Sec. \ref{}, at $x = 1/8$ the existence of SF solution of Eqs. (\ref{derivative of mathcalF A, lambda}) has not been  numerically confirmed (for \textit{var} method),  whereas   for the \textit{s-c} method the SF solutions of of Eqn.  (\ref{derivative of mathcalF lambda})  have been found  unstable against  Fermi sea (FS) with $\xi_2 =0$. In all four cases the SF$\to$FS transition is located at the critical concentrations $0.11 < x_c < 0.12$. However, our  numerical procedures for \textit{var} method  turned out to be unstable in the  vicinity of the $x_c$. For that reason, we chose doping $x = 13/128$ ($n \approx 0.898$),  which is in a safe distance from each of $x_c$, but for which the differences between the methods are   pronounced (as they generally increase with increasing doping).

The parameters of the Hamiltonian are the same as for the RVB case, except sign convention for $t$, now $t = 3 $ ($t = -3 $ in the RVB case), also $\Lambda_x = \Lambda_y = 512$. Again, we work  with low $T = 1/500$.
The thermodynamic potentials and mean-field   variables are listed in Tabs. III and IV. 

\begin{center}  
Table III. Equilibrium values of the thermodynamic potentials (per site) for SF solutions, for $n \approx 0.898 $, $\Lambda_{\tau} = 512$. $\tilde{\Omega}$ ($F$) stands for  $  \Omega -\lambda N$ ($\Omega + \mu N$)  for  \textit{var} and $\Omega_{s-c}$ ($\Omega_{s-c} +\mu_{s-c} N$) for \textit{s-c} methods, respectively.\\ 
\begin{tabular}{c c c c c}\hline \hline
Therm. Pot. & \textit{var} (I)  &  \textit{var} (II)   & \textit{s-c} (I)  & \textit{s-c}  (II)        \\  \hline 
$\Omega/\Lambda $  & -5.90103948  &   -5.75555848  &  -  &  - \\  
$\tilde{\Omega}/\Lambda$   &  -0.69003640     &  -0.71583672   &   -0.49006797   & -0.49344983  \\  
 $F/\Lambda $   &  -1.18762800  &   -1.16898668     &  -1.18536044   &  -1.16599431      \\  
\hline \hline
\end{tabular}
\end{center}
 
\begin{center}
Table IV. Values of chemical potentials and MF parameters for SF solutions for $n \approx 0.898$, $\Lambda_{\tau} = 512$.
$\tilde{\mu} $ stands for $\lambda + \mu $ (\textit{var}), and for $\mu_{s-c}$ (\textit{s-c}).\\
\begin{tabular}{c c c c c}\hline \hline

Variable & \textit{var} (I)  &  \textit{var} (II)   & \textit{s-c}  (I)  & \textit{s-c}  (II)   \\  \hline 
$\mu$  &    5.24623 &  5.10505  &  -    &  -   \\  
$\lambda $   &  -5.80007    &  -5.60943   &  -   & -  \\  
$\tilde{\mu}$   &  -0.55384     &  -0.50438     &   -0.77389      &  -0.74857      \\   
$\xi_1  $     &  0.19222     &  0.19321    & 0.18805      &    0.18844    \\  
$\xi_2$   &   0.10298       &  0.09840         &   0.11856    &   0.11731 \\
$ \eta_{1} $   &  -0.13476    &   -0.14848   &   -  &  -   \\ 
 $\eta_{2}$   &   -0.07220      &   -0.07562       &  - &  -  \\  
$ S_{AF} $   &   -0.23514       & -0.22522        & -0.24436    &  -0.23527    \\  
$\Phi_{\square} $   &  0.31311     &  0.29987    &   0.35811  &    0.35449  \\ 
$T_1 $   &  0.80695      &  0.77919     &   0.92799   &   0.91123  \\  
$T_2 $   &   0.18241      &  0.16008   & 0.29312  &    0.28009  \\  
\hline \hline
\end{tabular}
\end{center}
Within the standard mean-field approach (e.g. \cite{Marcin Polonica}) the (fictitious) flux  is defined as $\Phi_{\square} = \frac{1}{2\pi}\sum_{\langle ij \rangle \in \square} \text{Arg}(\chi_{ij})$, and $\square$ denotes plaquette composed from four bonds. Also,
for \textit{s-c} method, i.e. for $\eta_1 = \eta_2 = 0$, we have
\begin{equation} 
\text{Arg}(\chi_{ij}) = \arctan \left( \frac{\xi_2}{\xi_1}  \right) =  \arctan \left( \frac{T_2}{T_1 - t g^{t}}  \right),
\label{MF s-c flux II equal I}
\end{equation}
the last equality follows from (\ref{T tau 1 SF}). Interestingly, this equality holds  also for the variational approach with $\eta_1 \neq 0$,  $\eta_2 \neq 0$. It can be shown analytically, that $\xi_1/\xi_2 = \eta_1/\eta_2$, which together with (\ref{T tau 1 SF}) yields (\ref{MF s-c flux II equal I}).  Thus the two possible and \textit{a priori} different  definitions of $\Phi_{\square}$ within \textit{var} method turn out to be equivalent. Antiferromagnetic correlations are defined on the MF level as  $S_{AF} = -\frac{3}{2} g^{J}(\xi^2_1 + \xi^2_2)$, \cite{Didier, Marcin Polonica}.

\begin{figure}[h]
\begin{center}
\rotatebox{270}{\scalebox{0.45}{\includegraphics{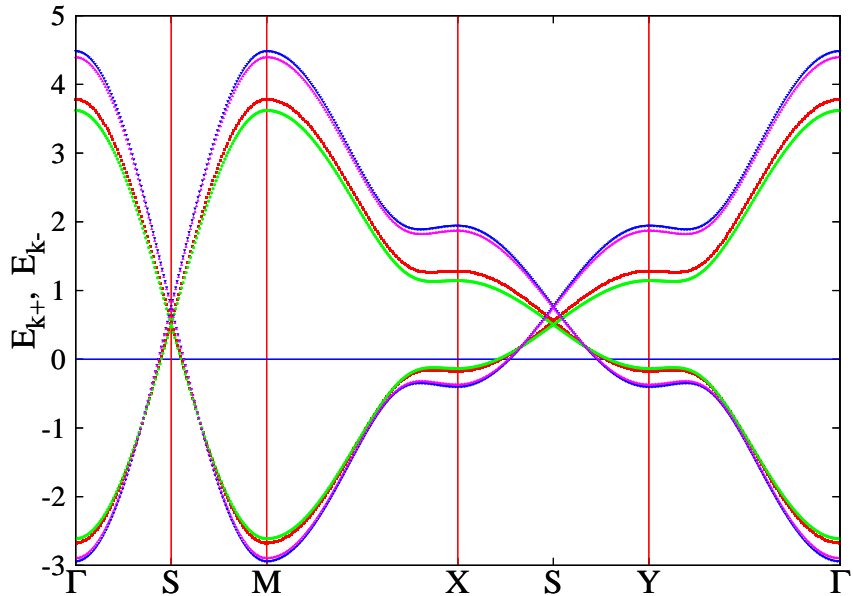}}}
\end{center}					
\caption{(Color online) Dispersion relations for both upper ($E_{\mathbf{k} +}$) and lower ($E_{\mathbf{k} -}$) subbands for the SF solutions  along the main symmetry lines in the Brillouin zone of the square lattice, of the size $\Lambda_x = \Lambda_y  = 512$, and for the filling $n\approx0.898$. Triangles -  self-consistent, non-variational  results for (I) and (II), red squares (I) and green circles (II)- the present variational method. Explicitly, for ($E_{\mathbf{k} +}$) near the maximum at point M,  from the bottom to the top: (II) \textit{var}, (I) \textit{var}, (II) \textit{s-c} (violet triangles), (I) \textit{s-c} (dark blue triangles).}
\label{dispertion for 512 SF}
\end{figure}

In contrast to the dRVB case, now the differences between renormalization schemes ((I) vs (II)) within each (\textit{var}, \textit{s-c}) method are small. This  is because in the absence of $\Delta_{ij}$, the $\chi_{ij}$ -dependences of   $g_{(I)}^{t}$  and  $g_{(II)}^{t}$ are quite similar, and the $\chi_{ij}$ -dependence of $g^{J}_{(II)}$ is weak for doping $x\approx 0.1$, thus causing no qualitative and only minor quantitative differences between  the renormalization schemes. On the other hand, the generic  modifications introduced by the variational approach within each RS are more significant.

Namely, from Tab. IV we see, that  \textit{s-c} method  favors SF more strongly than \textit{var} method,  which is indicated by the values of $\Phi_{\square}$ and $\xi_{2}$. Also,    $T_1$ and $T_2$, determining quasi-particle spectra, are smaller within \textit{var} method, and so are the quasi-particle energies, c.f. Fig. \ref{dispertion for 512 SF}.

\newpage

\section{Concluding remarks\label{summary}}
In summary, in this work we have compared, within two methods of approach, the two Gutzwiller renormalization schemes for the renormalized mean-field theory (RMFT) of t-J model in its simplest form. We emphasize  the advantages of the combination of the renormalization scheme  of Ref. \cite{Fukushima}  with the  variational method  proposed in Ref. \cite{JJJS_arx_0}.  First, a number of  theoretical arguments strongly favor this  choice. Moreover, in contrast to the other cases investigated by us (e.g. either non-variational method or renormalization scheme  of Ref. \cite{Sigrist}), the  former approach quite correctly predicts the upper critical doping $x_c \approx 0.27$ for a disappearance of the SC order. Also, the value of the optimal doping $x \approx 0.125$ is quite correctly predicted. In our opinion,  the formalism of Ref. \cite{Fukushima} augmented with the self-consistent variational treatment,  gives a chance for   the complete and  consistent one-particle  description  (in the form of RMFT) for a t-J model. Such   description, however, must encompass the t-J model   in its  complete  form \cite{Spalek Oles}, and include also more complicated symmetry breaking patterns. 
This is the subject of our current  investigation.

\subsection*{Appendix A: Equivalence relation for  mean-field Hamiltonians \label{AppendixA}}
Below we present some details of our    formalism, which are  necessary for the present disscusion.  We  also comment on the relationship between our method and the formalisms of   Ref. \cite{Kai}.  We start from the MF Hamiltonian   of the  form
\begin{equation}
\hat{H}(\vec{A}) = \hat{H}_{e0} +  C_{0}(\vec{A})\cdot\hat{\mathbf{1}}_{D} + \sum_{s=1}^{M}   C_{s}(\vec{A})\hat{A}_{s} + \sum_{w=1}^{M^{\prime}}   G_{w}(\vec{A})\hat{B}_{w}. 
\label{MF H specyfic}
\end{equation}
In the above,   $\hat{H}_{e0}$ is an  $\vec{A}$-independent part,   $C_{0}(\vec{A})$, $C_{s}(\vec{A})$ and $G_{w}(\vec{A})$  are some complex-valued functions of mean-fields $\vec{A}$.  Operators  $\hat{B}_{w}$ are those, which average values are not present in $\hat{H}(\vec{A})$.  We also assume that  all the  operators appearing above are bilinear in creation and/or annihilation operators. From (\ref{MF H specyfic})  we have
\begin{eqnarray}
\hat{H}_{\lambda}(\vec{A}) &=&  \hat{H}(\vec{A})-\sum_{s=1}^{M}   \lambda_{s}(\hat{A}_{s} - A_{s} ) \nonumber \\  &=& \hat{H}_{e0} + C^{\lambda}_{0}(\vec{A})\cdot\hat{\mathbf{1}}_{D} +  \sum_{s=1}^{M} C^{\lambda}_{s}(\vec{A})\hat{A}_{s} + \sum_{w=1}^{M^{\prime}}   G_{w}(\vec{A})\hat{B}_{w},  
\label{ren H with lambda gen not}
\end{eqnarray}
with $C^{\lambda}_{0} = C_{0} +  \sum_{s=1}^{M} \lambda_{s} A_{s} $,
$C^{\lambda}_{s} = C_{s}  - \lambda_{s} $. For a given value of $\vec{A}$, the self-consistency equations (second half of the $2M$ equations (\ref{derivative of mathcalF A, lambda})) may be written as 
\begin{eqnarray}
A_{t} &\equiv& \langle \hat{A}_{t} \rangle_{\lambda} = \text{Tr}[\hat{A}_{t}\hat{\rho}_{\lambda}] \nonumber \\ &=&\frac{\text{Tr}[\hat{A}_{t} \exp\big(-\beta (\hat{K}_{e0} + \sum_{s=1}^{M}   C^{\lambda}_{s}(\vec{A})\hat{A}_{s} + \sum_{w=1}^{M^{\prime}}   G_{w}(\vec{A})\hat{B}_{w} )\big)]}{\text{Tr}[ \exp\big(-\beta (\hat{K}_{e0} + \sum_{s=1}^{M}   C^{\lambda}_{s}(\vec{A})\hat{A}_{s}+ \sum_{w=1}^{M^{\prime}}   G_{w}(\vec{A})\hat{B}_{w}\big)]}.
\label{The return of the self consistent equations}
\end{eqnarray} 
In above, the $C^{\lambda}_{0}(\vec{A})$ term canceled, and $\hat{K}_{e0} = \hat{H}_{e0} - \mu \hat{N}$.   
Eqs. (\ref{The return of the self consistent equations}) may be formally solved  (which usually cannot be achieved in an analytic fashion) for $\vec{\lambda}$, and then  $\vec{\lambda} = \vec{\lambda}(\vec{A})$, thus for  a given $\vec{A}$, Eqs. (\ref{The return of the self consistent equations}) determine\footnote{We assume that the solution exists and is unique, or there exist a finite number of the solutions, but one of them may be  unambiguously selected.} $C^{\lambda}_{s}(\vec{A})$. As a result, the $\vec{A}$-dependence of the  coefficients $C^{\lambda}_{s}(\vec{A})$ is determined by the choice of operators $\hat{H}_{e0}$, $\{\hat{A}_{s}\}^{M}_{s=1}$ and  $\{\hat{B}_{w}\}^{M^{\prime}}_{w=1}$. If we change the original form of the coupling of $\{\hat{A}_{s}\}^{M}_{s=1}$ operators  to the mean fields, according to
\begin{equation}
C_{s}(\vec{A}) \to \tilde{C}_{s}(\vec{A}) = \varphi_s(C_{s}(\vec{A})),
\label{redefinition of Couplings}
\end{equation}
but without changing  $G_{w}(\vec{A})$ functions, then  the  Lagrange multipliers change in a way that   $C^{\lambda}_{s}(\vec{A})$ is unchanged, in order to fulfill (\ref{The return of the self consistent equations}). The only $\vec{A}$-dependent part which may be non-trivially modified by (\ref{redefinition of Couplings}) is $C_{0}(\vec{A})\cdot\hat{\mathbf{1}}_{D}$. Obviously, $\mathcal{S}_{\lambda}$, defined as (c.f. \cite{JJJS_arx_0})
\begin{equation}
\mathcal{S}_{\lambda} =  \text{Tr} \Big[ - \hat{\rho}_{\lambda} \ln  \hat{\rho}_{\lambda} -\beta \big( \hat{\rho}_{\lambda} \hat{H}
-  \mu \hat{\rho}_{\lambda} \hat{N} - \sum_{s=1}^{M}  \lambda_{s}\hat{\rho}_{\lambda} (\hat{A}_{s} -  A_{s})\big) - \omega (\hat{\rho}_{\lambda} - \frac{1}{D})\Big]
\label{entropy MF}
\end{equation}
and $\mathcal{F}(\vec{A}, \vec{\lambda}) \equiv  -\beta^{-1} \mathcal{S}_{\lambda}( \hat{\rho}_{\lambda}(\vec{A}, \vec{\lambda}), \vec{A}, \vec{\lambda})$
may be also modified. However, the density operator $\hat{\rho}_{\lambda}$ is invariant under (\ref{redefinition of Couplings}), so are, for given $\vec{A}$,  all the averages, also those of $\hat{H}_{e0}$ and $\hat{B}_{w}$ operators. Moreover, if the transformations (\ref{redefinition of Couplings}) are such   that $\langle \hat{H}_{\lambda} \rangle_{\lambda} =   \langle \hat{H} \rangle_{\lambda}$ remains unchanged, the value of  $\mathcal{F}(\vec{A}, \vec{\lambda}({\vec{A}})) $  is   not modified. Consequently, in such a situation, the equilibrium values of mean fields, as well as of the quantities $C^{\lambda}_{s}(\vec{A})$ (e.g. $T_{\tau}, D_{\tau}$ of Eqn. (\ref{T tau}) or $T_{1}, T_{2}$ of Eqn. (\ref{T tau SF})) are also invariants  of (\ref{redefinition of Couplings}). Summarizing, all MF Hamiltonians  constructed from  the same set  of operators $\hat{H}_{e0}$, $\{\hat{A}_{s}\}^{M}_{s=1}$ and  $\{\hat{B}_{w}\}^{M^{\prime}}_{w=1}$   and having the same $\vec{A}$-dependence of the expectation value,  are   equivalent, and belong to the same  equivalence class. The transformations (\ref{redefinition of Couplings}) may be viewed in   analogy to gauge transformations, not changing the physical content of the model.   

Suppose, that  the MF Hamiltonian (\ref{MF H specyfic}) is such that
\begin{equation}
 \hat{H}_{e0} = 0, ~~~~~~~~\forall_w:~~~~ G_{w} = 0.
\label{complete H}
\end{equation}
Then  the equilibrium values of  $\lambda_{s}$ may be easily obtained in an analytic  fashion, using Eqn. (17) of Ref. \cite{JJJS_arx_0}, i.e.

\begin{equation}
    \lambda_w     =  - \Big \langle  \frac{\partial  \hat{H} }{\partial A_w} \Big \rangle_{\lambda}.  
\label{mean value of H GMF with lambda, d/dA}
\end{equation} 
Now, we choose a transformation  (\ref{redefinition of Couplings}) of a  specific form, 
\begin{equation}
C_{s}(\vec{A}) \to \tilde{C}_{s}(\vec{A}) = 0, ~~~ s \neq 0 \nonumber
\label{redefinition of Couplings II}
\end{equation}
\begin{equation}
C_{0}(\vec{A}) \to \tilde{C}_{0}(\vec{A}) = C_{0}(\vec{A}) + \sum_{s} C_{s}(\vec{A}) A_s.  
\label{redefinition of Couplings III}
\end{equation}
This yields 
 \begin{equation}
\hat{H}^{(\sim)}_{\lambda}(\vec{A}) =  - \sum_{s=1}^{M}  \lambda_{s}  (\hat{A}_{s} -  A_{s}) + W(\vec{A}),
\label{simples H lambda}
\end{equation}
with $  \hat{H}(\vec{A}) = W(\vec{A})\hat{\mathbf{1}}_{D} $,   $W(\vec{A}) = \langle \hat{H}(\vec{A}) \rangle = \tilde{C}_{0}(\vec{A})$.  Using (\ref{mean value of H GMF with lambda, d/dA}) we may rewrite    (\ref{simples H lambda}) as
 \begin{equation}
\hat{H}^{(\sim)}_{\lambda}(\vec{A}) =    \sum_{s=1}^{M}  \frac{\partial W(\vec{A})}{\partial A_s}  (\hat{A}_{s} -  A_{s})  + W(\vec{A}),
\label{simples H lambda wyrazony eksplicite}
\end{equation}
because $ \partial W(\vec{A})/\partial A_s = \partial \langle \hat{H}(\vec{A}) \rangle/\partial A_s = \langle\partial  \hat{H}(\vec{A}) /\partial A_s \rangle $.
The form (\ref{simples H lambda wyrazony eksplicite}) is the most convenient, as half of the variables ($ \vec{\lambda}$) are eliminated, which reduces the number of equations to be solved numerically.

 Also, in the limit  ($\beta \to \infty$) the results of   finite-temperature formalism are essentially identical  to those the true $T=0$ analysis, and we may compare them with those of Ref. \cite{Kai}.
Hamiltonian of that Reference  reads in our notation 
 \begin{equation}
\hat{H}^{(K)}_{\lambda}(\vec{A}) =  - \sum_{s=1}^{M}  \lambda_{s}(\vec{A})   \hat{A}_{s} =  \sum_{s=1}^{M}   \frac{\partial W(\vec{A})}{\partial A_s}   \hat{A}_{s},
\label{simples H lambda Kai}
\end{equation}
and differs from  (\ref{simples H lambda}) only by the constant term  $ \sum_{s=1}^{M}  \lambda_{s} A_{s}  +  \langle \hat{H}(\vec{A}) \rangle$. Consequently, both (\ref{simples H lambda}) and (\ref{simples H lambda Kai}) have the same eigenvalues and eigenvectors (the presence of $-\mu \hat{N}$ obviously does not change the above arguments).
Please note, that the analytical evaluation of Lagrange multipliers through (\ref{mean value of H GMF with lambda, d/dA}), and hence the  application of the method of Ref. \cite{Kai}  is possible only if the conditions (\ref{complete H}) are fulfilled, but not for the general form (\ref{MF H specyfic}) of the MF Hamiltonian.

\subsection*{Acknowledgments}
An  invaluable technical help that one of the authors (JJ) received    from Andrzej Kapanowski, Jan Kaczmarczyk, Gosia Kaliszan and Micha\l~K\l os  is   warmly acknowledged.
All the numerical computation was performed using GSL (Gnu Scientific Library) efficient procedures. 
The authors acknowledge the Grant from Ministry of Science and Higher Education.

\vspace{0.3cm}
$\ast$ e-mail: jedrak@th.if.uj.edu.pl \\

$\dag$ e-mail: ufspalek@if.uj.edu.pl

\end{document}